\documentclass[final,5p,times,twocolumn]{elsarticle}

\usepackage{amssymb}
\usepackage{amsmath}
\usepackage{amsfonts}
\usepackage{amssymb}
\usepackage{calligra}
\usepackage{bm}
\usepackage{graphicx}
\usepackage{float}
\usepackage{subfigure}
\usepackage{stackengine}
\usepackage{natbib}
\usepackage{hyperref}
\usepackage{physics}

\journal{Physica A: Statistical Mechanics and its Applications}

\begin{document}

\begin{frontmatter}
\author{Mohammad Pouranvari}
\ead{m.pouranvari@umz.ac.ir}
\affiliation{organization={Department of Solid-State Physics, Faculty of Science, University of Mazandaran},
city={Babolsar},
postcode={4741613534},
country={Iran}}

\title{Entanglement entropy of XX spin $1/2$ chain with random
	partitioning at arbitrary temperature}

\begin{abstract}
  We study the entanglement properties of random XX spin $1/2$ chains
  at an arbitrary temperature $T$ using random partitioning, where
  sites of a size-varying subsystem are chosen randomly with a uniform
  probability $p$, and then an average over subsystem possibilities is
  taken. We show analytically and numerically, using the approximate
  method of real space renormalization group, that random partitioning
  entanglement entropy for the XX spin chain of size $L$ behaves like
  EE$(T,p) = a(T,p) L$ at an arbitrary temperature $T$ with a uniform
  probability $p$, i.e., it obeys volume law. We demonstrate that
  $a(T,p) = \ln(2) \langle P_s + P_{t_{\uparrow\downarrow}} \rangle
  p(1-p)$, where $P_s$ and $P_{t_{\uparrow\downarrow}}$ are the
  average probabilities of having singlet and
  triplet$_{\uparrow\downarrow}$ in the entire system,
  respectively. We also study the temperature dependence of pre-factor
  $a(T,p)$. We show that EE with random partitioning reveals both
  short- and long-range correlations in the entire system.
\end{abstract}

\end{frontmatter}

\section{Introduction}\label{sec:introduction}
Physicists use the entanglement properties of a system (among other
characterizations) to understand the system, theoretically and
experimentally\cite{RevModPhys.81.865,LAFLORENCIE20161,Kauffman_2002,
  PhysRevX.7.021021,PhysRevLett.96.110404,Calabrese_2004}. The notion
of entanglement was born with quantum physics\cite{PhysRev.47.777,
  schrodinger_1935}, and it has no corresponding concept in classical
physics. The entanglement properties capture the non-local properties
of the system related to the correlations in the entire system. There
are some measures to quantify the entanglement properties of the
system, among which we can name the entanglement entropy (EE) (see
below for definition). However, there are other measurements such as
Renyi entropy, concurrence, logarithmic negativity
etc.\cite{RevModPhys.81.865,PhysRevA.92.042329,PhysRevLett.78.2275}. It
is also possible to measure the entanglement properties of the system
experimentally\cite{RevModPhys.73.565,PhysRevA.99.062309,Sackett2000}. One
of the features that people study is the behavior of the EE versus
system size.  For example, the EE grows with the boundary of the
subsystem (what is called area law) or it grows with the volume of the
subsystem (volume law). For free fermions, the area law sometimes is
violated\cite{RevModPhys.82.277,
  PhysRevLett.96.010404,Vitagliano_2010,
  PhysRevB.89.115104,PhysRevLett.109.267203,PhysRevLett.105.050502}.
To calculate the EE, one \emph{usually} divides the system into two
parts. If the system has $L$ sites, the first $L/2$ sites are the
subsystem (sites $1$ up to $L/2$), and the rest is called the
environment. In this cutting, the amount of entanglement between this
specific subsystem and its environment is calculated. Studying the EE
with this kind of bi-partitioning has been used before to analyze the
behavior of different systems at zero temperatures or a non-zero
temperature\cite{Wong2013, Alba_2009, PhysRevB.90.220202,
  PhysRevLett.110.091602, PhysRevB.100.165135,Rodriguez-Laguna_2017,
  PhysRevB.101.205121, Ramirez_2014}. Rather than simply cutting the
system in the middle, there are also other cutting possibilities that
affect the information we can obtain. I.e., for a system with a phase
transition, the EE obtained with one form of cutting could reveal the
phase transition, while we get no information about the phase
transition with some other cuttings. Thus, \emph{the way we cut the
  system matters}. See, for example,
Refs. \cite{ahmadi2021frustrated,Moradi_2016,PhysRevB.101.195117}
where people examined different cuttings in bi-partitioning. Thus, to
acquire all the entanglement properties of the system, we would divide
it in all possible ways. In this regard, we use the notion of
\emph{random partitioning}, in which we take an average over the EE's
corresponding to the randomly chosen subsystems. More accurately, to
obtain random-partitioning EE, we do the following: first, we
attribute a probability to each site based on a probability
distribution, i.e., sites that belong to the subsystem are chosen
randomly. Moreover, subsystem size varies between $1$ to $L$ (there
are $n \choose L$ different forms of having subsystems with $n$
sites). We consider all of these ways of partitioning, and we take an
average of the EE corresponding to them. This way of partitioning is
entirely different from the usual method of bi-partitioning. In the
bi-partitioning, the system's middle is the boundary between the
subsystem and its environment. In contrast, in the random
bi-partitioning, there are many boundaries at different points. As a
limiting case, if the subsystem sites are every other spin, we expect
a volume law for the entanglement entropy.

Recently, this form of partitioning has been used for the \emph{ground
  state} of a \emph{clean}, free fermion system\cite{Roosz2020}. They
found a volume law dependence of the EE to the system size with a
logarithmic correction term.  Also, people studied the entanglement
spectrum under random partitioning\cite{PhysRevB.91.220101}. Here, we
advance these previous studies in two directions. First, we consider a
\textit{disordered} system with random impurities and second, we are
concerned mainly with a \textit{typical excited state} at an arbitrary
temperature $T$.

By analytical and numerical investigations, we find that the behavior
of the EE with a random partitioning is volume law:
$EE(T,p) = a(T,p)L$, with a pre-factor $a$ that depends on the
probability and temperature. We obtain the analytical form for
$a(T,p)$ and verify it numerically. We show that the behavior of
$a(T,p)$ is related to the number of entangled bonds \emph{in the
  entire system}. We finally discuss that EE in the random
partitioning reveals the short-range and the long-range correlations
in the entire system.

The rest of the paper is separated into two parts; in the first part,
section \ref{sec:model-method}, we introduce the XX model and the
method to calculate the EE. The calculations are based on the
approximate method of real space renormalization group for the ground
state (RSRG) and a typical excited state (RSRG-X). We explain these
methods in the following section. We then analytically prove the
behavior of the EE in random partitioning with a uniform
probability. The numerical evaluations and verification of the
behavior of the EE are presented in section \ref{sec:num}. Based on
these calculations, we give a picture of the EE in random partitioning
and conclude in section \ref{sec:conclusion}.

\section{Analytical evaluation}\label{sec:model-method}

We consider a one-dimensional XX spin $1/2$ chain with $L$ spins. They
are coupled together locally and with random strength. Hamiltonian of
the system is:
\begin{equation} \label{eq:H}
	H=\sum_{n=1} ^{L-1} J_{n} (s_{n}^{x} s_{n+1}^{x}+s_{n}^{y}s_{n+1}^{y}) .
\end{equation}
(with open boundary condition) where $J_n$ are distributed randomly by
a distribution function (to be determined later). The entanglement
properties of this model have been studied
before\cite{PhysRevB.83.045110, Refael_2009,
  mohdeb2022excitedeigenstate, PhysRevB.72.140408}.

We want to obtain the entanglement properties of this model, and to do
so, we use the notion of EE. In a \emph{bi-partitioned system}, where the
system is divided into two parts: subsystem $A$ and its environment,
the EE is the von Neumann entropy of the reduced density matrix of the
subsystem:
\begin{equation} \label{EE}
	\text{EE}=-Tr [\rho_A \ln \rho_A],
\end{equation}
where $\rho_A$ is the reduced density matrix for the subsystem $A$
obtained by tracing out the environment degrees of freedom from the
density matrix $\rho$ of the entire system. For a pure state the
density matrix is $\rho = \ket{\Psi}\bra{\Psi}$, where $\ket{\Psi}$ is
the state of the system. Thus, to calculate the EE in a bi-partitioned
system in a brute-force method, first, we need to diagonalize the
Hamiltonian to obtain its eigenvalues and eigenvectors
$\{\Psi\}$. Then, by calculating the reduced density matrix of a
chosen subsystem and using Eq. (\ref{EE}), we obtain the EE.  To
obtain the system's state in the XX spin $1/2$ chain with $L$ spins,
one deals with matrices with size $2^L$, which exponentially growing
with the system size. By Jordan-Wigner transformation, the XX model is
transformed into a free fermion model, where the size of the matrices
to deal with in numerical calculations reduces to
$L$\cite{0305-4470-36-14-101, PhysRevB.69.075111}. In addition to this
direct and exact method, the XX spin $1/2$ chain has been studied by
approximate methods such as the real space renormalization group
(RSRG) method\cite{PhysRevB.22.1305, PhysRevB.102.014455,
  Igloi2018}. In this method, the \emph{approximate} ground state of a
\emph{random} XX spin $1/2$ chain is obtained. An extension of this
method is also developed to obtain an approximate typical \emph{
  excited state}, namely the RSRG-X. We use these approximate methods
to calculate the EE in random partitioning.

We should note that, in this paper, we do not use the corresponding
free fermion model of Eq. (\ref{eq:H}) to obtain the EE in random
partitioning. Using the free fermion method in calculating EE for
disjoint blocks is incorrect (since the Jordan-Wigner transformation
of $i$th site depends on the previous sites). Instead, we use
RSRG/RSRG-X approximate method to obtain EE, which is based on the
number of singlets and triplets that cross the boundary of the
subsystem. These methods do not fail for a disjoint subsystem. In what
follows, we explain both methods and describe how to use them to
calculate approximately the EE.

\subsection{Real space renormalization group} 
\begin{figure}
	\includegraphics[width=\linewidth]{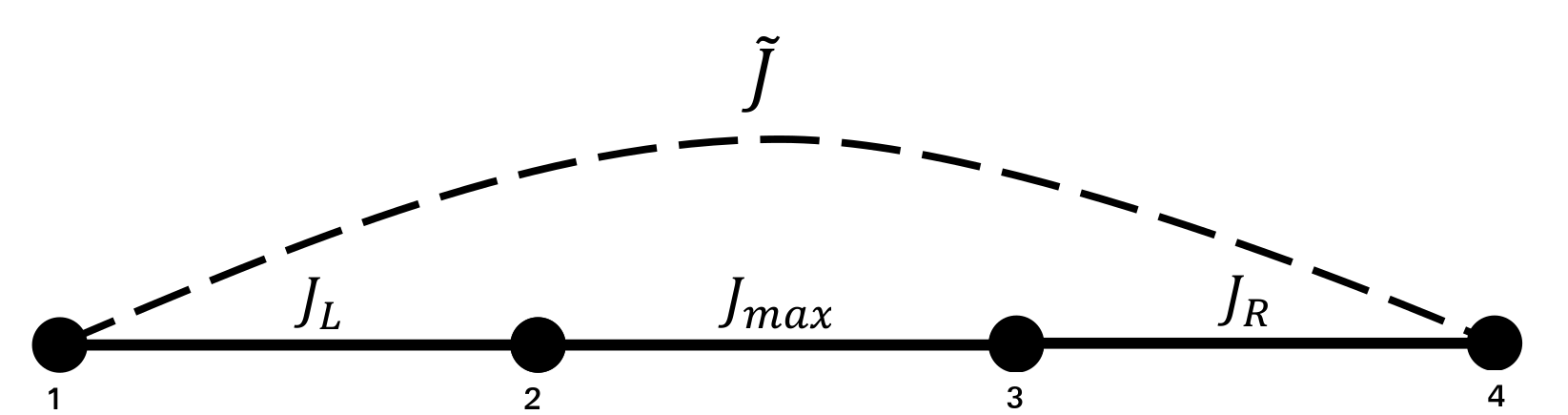}
	\caption{\label{rsrg} A part of the spin chain is presented in
          which we assume $J_{max}>J_L,J_R$. In real space
          renormalization group method, after picking up $J_{max}$, we
          remove spin $2$ and $3$ and couple spin number $1$ and $4$
          with $\tilde{J}=\frac{J_L J_R}{J_{max}}$.}
\end{figure}

{\label{sec:rsrg}}
To obtain the approximate ground state of a XX spin $1/2$ chain by RSRG
method, we go through the following
steps\cite{PhysRevB.50.3799}. Consider the Hamiltonian of
Eq. (\ref{eq:H}) in which coupling constants $\{J\}$ are distributed
randomly. First, we pick up the largest coupling $J_{max}$. In the
ground state of the system, we put the two spins that are coupled with
$J_{max}$ in a \emph{singlet state} $\ket{\text{singlet}} =
\frac{1}{\sqrt{2}}[\ket{\uparrow\downarrow} -
  \ket{\downarrow\uparrow}]$. Next, we remove these two spins and
couple the two closest spins with an effective coupling
$\tilde{J}=\frac{J_L J_R}{J_{max}}$ (see Fig.\ref{rsrg} for a
schematic representation).

By repeating this process, the selected spin pairs are in the singlet
state. Thus, the ground state of the system is the direct product of
singlets with arbitrary bond length (the so-called random singlet
phase). Since $\tilde{J}$ is smaller than both of $J_L, J_R$, by
repeating the RSRG process, we will get smaller and smaller values for
$\tilde{J}$. So, the probability of $\{J\}$ distributions will be a
power-law: $P(J) = \alpha \ J^{\alpha-1} \text{ for } 0 \leq J \leq
1$. The RSRG fixed point is the infinite randomness fixed point and
corresponds to $\alpha \to 0$. In this regard, small values of
$\alpha$ correspond to the strong disorder regime that RSRG yields to
an asymptotically correct ground state.

For a highly excited state, a modified version of the RSRG is developed,
namely the RSRG-X\cite{PhysRevX.4.011052}. We should note that, at
$T=0$, the state of the system is pure and we obtained it
approximately using the RSRG method, which in this approximation is
the product state of the singlet state pairs. On the other hand, for $T\ne
0$, the state of the system is mixed which represents a combination of
possible states weighted with a probability. In this paper, we do not
consider a mixed state for a non-zero temperature, but we consider a
typical state which is one of the possible states. This typical state
(which in the RSRG-X approximation is a product state of singlet and
each of the triplet states) is a pure one and the notion of the EE can
be used to quantify the entanglement properties in the system.

In the RSRG-X method, we look for the two spins that are coupled with
the largest magnitude value, and put them in the singlet state or each
of the triplet states based on the Boltzmann distribution function:
\begin{equation}\label{eq:Boltzmann}
	P_B=\frac{1}{Z} \exp(-E/T),
\end{equation}
where $E$ is the energy of the singlet/triplet state of the two spins,
and $Z=2+2 \cosh{\frac{J}{2T}}$ to have a normalized probability (see
Table \ref{tb:eigen}\cite{PhysRevB.90.220202}).

\begin{table}
  \caption{\label{tb:eigen} The effective coupling based on the chosen
    eigen-state of two spins and the corresponding energy.}
  \begin{tabular}{l|c|c|c}
    Eigen-state & Eigenvalue & Probability & \vtop{\hbox{\strut Effective}\hbox{\strut coupling}}   \\
    \hline
    singlet= $\frac{1}{\sqrt{2}}[\ket{\uparrow\downarrow} - \ket{\downarrow\uparrow}]$ & $-J/2$ & $\frac{1}{Z} e^{J/2T}$ & $\tilde{J} \approx +\frac{J_L J_R}{J_{max}}$ \\
    triplet$_{\uparrow\downarrow}$=$\frac{1}{\sqrt{2}}[\ket{\uparrow\downarrow} + \ket{\downarrow\uparrow}]$ & $+J/2$ & $ \frac{1}{Z} e^{-J/2T}$ &$\tilde{J} \approx +\frac{J_L J_R}{J_{max}}$ \\
    triplet$_{\uparrow\uparrow}$= $\ket{\uparrow\uparrow}$ & $0$ & $\frac{1}{Z}$ & $\tilde{J} \approx -\frac{J_L J_R}{J_{max}}$ \\
    triplet$_{\downarrow\downarrow}$= $\ket{\downarrow\downarrow}$ & $0$ & $\frac{1}{Z}$ & $\tilde{J} \approx -\frac{J_L J_R}{J_{max}}$ \\

  \end{tabular}
\end{table}

The effective coupling, $\tilde{J}$ depends on which singlet or
triplets are chosen by the Boltzmann distribution. In doing the RSRG-X
method, the probability of getting smaller magnitude values for
couplings increases, and thus we have a power-law distribution:
\begin{equation}\label{eq:PJ}
	P(J) = \frac{\alpha}{2} |J|^{\alpha-1} \text{,\  for } |J| \leq 1
\end{equation}
Like the RSRG method, the fixed point corresponds to $\alpha \to 0$
and the strong disorder regime, where the RSRG-X is asymptotically
correct, corresponds to small values of $\alpha$.  The outcome of the
RSRG-X method, a typical excited state, is the direct product
of singlets and triplets. We work in the sector of half-filling in the
corresponding free fermion representation, which is equivalent to
$S_z^{total}=0$ in the spin representation of the Hamiltonian.

We note that the singlet and triplet$_{\uparrow\downarrow }$ states
are entangled states with the value of the EE equal to $\ln(2)$; but
triplet$_{\uparrow\uparrow}$ and triplet$_{\downarrow\downarrow}$
states are not entangled. Since the ground state (a typical excited
state) of the system in the RSRG (RSRG-X) method is the product state
of singlets (singlets and triplets), only those singlets (singlets and
triplet$_{\uparrow\downarrow }$) that \emph{cross the boundary}
contribute to the EE.  Thus, for a bi-partitioned system, to calculate
the EE, we count the number of singlets (number of singlets and
triplet$_{\uparrow\downarrow}$) crossing the boundary and multiply it
by $\ln(2)$\cite{PhysRevLett.93.260602, Refael_2009}. The numerical
verification of the RSRG and RSRG-X methods to calculate the EE have
been studied before\cite{PhysRevB.88.075123, PhysRevB.92.245134,
  PhysRevB.72.140408, PhysRevB.90.220202}.

At $T=0$, all spins that are decimated in the RSRG method are in the
singlet state. Namely, the probability of having a singlet, $P_s$ is
$1$, and the probability of having each of the triplets is $0$. Thus
we expect $L/2$ singlets across the entire system.  On the other hand,
we expect that each of the singlet and the triplets are chosen with
the same probability for a large $T$ in the RSRG-X method. I.e.,
$P_s = P_{t_{\uparrow\downarrow}}= P_{t_{\uparrow\uparrow}} =
P_{t_{\downarrow\downarrow}}= \frac{1}{4}$. See Table
\ref{tb:eigen}. As $T \to \infty$, the Boltzmann probability of the
singlet and triplets are the same. Therefore, there are $L/4$ singlets
and $L/4$ triplet$_{\uparrow\downarrow}$s. In addition, for an
arbitrary $T \ne 0$, we need to calculate the average probabilities
weighted with $P(J)$:

\begin{eqnarray}
	\langle P_s \rangle &=& \int_{-1}^{1} \dd{J} P(J)
        \frac{e^{J/2T}}{Z}, \\ \langle P_{t_{\uparrow\downarrow}}
        \rangle &=& \int_{-1}^{1} \dd{J} P(J) \frac{e^{-J/2T}}{Z},
\end{eqnarray}
and thus, the average probability of having a singlet and
a triplet$_{\uparrow\downarrow}$ at an arbitrary temperature $T$ is:
\begin{equation}
	\langle P_s + P_{t_{\uparrow\downarrow}} \rangle = \alpha (2T)^{\alpha} \int_{0}^{\frac{1}{2T}} \dd{x} \frac{x^{\alpha-1}}{1+\sech(x)}.
\end{equation}
There is no simple analytical solution for this integral, so we will
calculate it numerically. Finally, we should note that the RSRG is not
an exact method, and it is asymptotically
correct\cite{PhysRevB.88.075123}.

\subsection{EE  in random partitioning}\label{sec:EE_RP}
Now, we explain how to calculate the EE for a bi-partitioned system in
which the sites that belong to the subsystem are chosen
\emph{randomly}. First, we specify a probability $p_i$ for each site
$i$ to belong to the subsystem based on a probability
distribution. The subsystem size $n$, can vary from $1$ to $L$, and
for each of them, there are $L \choose n$ different ways of choosing
$n$ sites out of $L$ sites. In addition, for each of these choices,
there is a corresponding probability that each of the $n$ sites
belongs to the subsystem and other sites do not belong to the
subsystem: $\prod_{i \in A} p_i \prod_{i \notin A} (1-p_i)$.  Thus the
EE for a specific probability distribution in the random partitioning
$\text{EE}(T, \{p\})$ is the following\cite{Roosz2020}:

\begin{equation}\label{eq:direct}
  \text{EE}(T, \{p\}) = \sum_{n=1}^{L-1}  \sum_{j=1}^{{L \choose n}}   EE_{n}^j (T) \left( \prod_{i \in A} p_i \right ) \left( \prod_{i \notin A} (1-p_i)\right) ,
\end{equation}
where $EE_{n}^j$ is one of the $ L \choose n$ calculated EE's
corresponding to the case of having $n$ sites in the subsystem.

In the particular case of uniform probability, where the probability
for each site to belong to the subsystem is the same for all sites
($p_i = \text{constant}=p$) we can use the above equation to obtain
the EE in random partitioning with constant probability distribution
EE$(T,p)$:

\begin{equation}
  \text{EE}(T, p)  =  \sum_{n=1}^{L-1}  \sum_{j=1}^{{L \choose n} } EE_{n}^j(T) \ p^{n} (1-p)^{L-n}.
\end{equation}

In practice, we can not go over all samples of ${L \choose n}$ choices
for a large $L$; instead, we take the average over enough large number
of samples to obtain $\overline{EE}_{n}$, and thus we have:

\begin{equation}\label{eq:EEp}
  \text{EE}(T, p) = \sum_{n=1}^{L-1} \overline{EE}_{n}(T)  {L \choose n}  p^{n} (1-p)^{L-n},
\end{equation}
Since ${L \choose n}={L \choose L-n}$, we can deduce from
Eq. (\ref{eq:EEp}) that $\text{EE}(T, p) = \text{EE}(T, 1-p)$, and
thus we would expect a symmetric plot for EE$(p)$ versus $p$ about
$p=1/2$.

To obtain an analytical expression for $\overline{\text{EE}}_{n}(T)$,
we do the following: at $T=0$, since the state of the system is pure,
we expect that $\overline{\text{EE}}_{n} =\overline{\text{EE}}_{L-n}$;
thus, we guess that we can write
$\overline{\text{EE}}_{n} \propto n(L-n)$. To obtain the
proportionality at zero temperature, we note that all bonds are
singlet; therefore, having one site as the subsystem will yield to
$\overline{\text{EE}}_{n=1} = \ln(2) \times 1$; thus the
proportionality is $\frac{\ln(2)}{L-1}$, and:
\begin{equation}\label{EET0}
  \overline{\text{EE}}_{n}(T=0) = \frac{\ln(2)}{L-1} n(L-n).
\end{equation}

Replacing this result in Eq. (\ref{eq:EEp}), we obtain the following
expression for the EE$(p)$ at $T=0$:
\begin{equation}\label{eq:EEt0p}
	\text{EE}(T=0,p) = \ln(2) p(1-p) L.
\end{equation}

In addition, for an arbitrary $T$, as we argued above, on average,
only $\langle P_s + P_{t_{\uparrow\downarrow}} \rangle$ fraction of
the bonds contribute to the EE and thus:

\begin{eqnarray}
  \text{EE}(T, p) &=& \ln(2) \langle P_s + P_{t_{\uparrow\downarrow}} \rangle  p(1-p) L  \label{EEpt} \\
                  &=& a(T,p) \ L \label{EEpt1}
\end{eqnarray}
where,
$a(T,p) = \ln(2) \langle P_s + P_{t_{\uparrow\downarrow}} \rangle \
p(1-p)$. In conclusion, we see a volume law expression for the random
partitioning EE with $a(T,p)$ as the pre-factor as a function of
temperature and probability. We present numerical calculations in the
next section that verify our analytical results.

\section{Numerical verification}\label{sec:num}
In our numerical calculations of the EE in a random partitioning with
a uniform probability, we do the following. First, we apply the
RSRG/RSRG-X to obtain the approximate ground state/typical excited
state of the system corresponding to a specific system size. To work
in the strong disorder regime where the RSRG method is asymptotically
correct, we set $\alpha=0.2$. Then we randomly choose sites that
belong to the subsystem. By counting the number of singlets and
triplet$_{\uparrow\downarrow}$s that cross the boundary of the chosen
subsystem, we calculate EE for that chosen subsystem. We repeat this
process large enough times and calculate its average
$\overline{\text{EE}}_n$.  In the RSRG-X process, since each singlet
or triplet is chosen based on the Boltzmann distribution, we need to
take the ensemble average for each temperature $T$. In addition, since
the coupling constants are random, we also take the disorder average
over random $\{J\}$ realizations. After doing these averaging
calculations, the EE$(T, p)$ is obtained.

\begin{figure*}
  \centering
  \begin{subfigure}{}%
    \def\big{\includegraphics[width=0.31\textwidth]{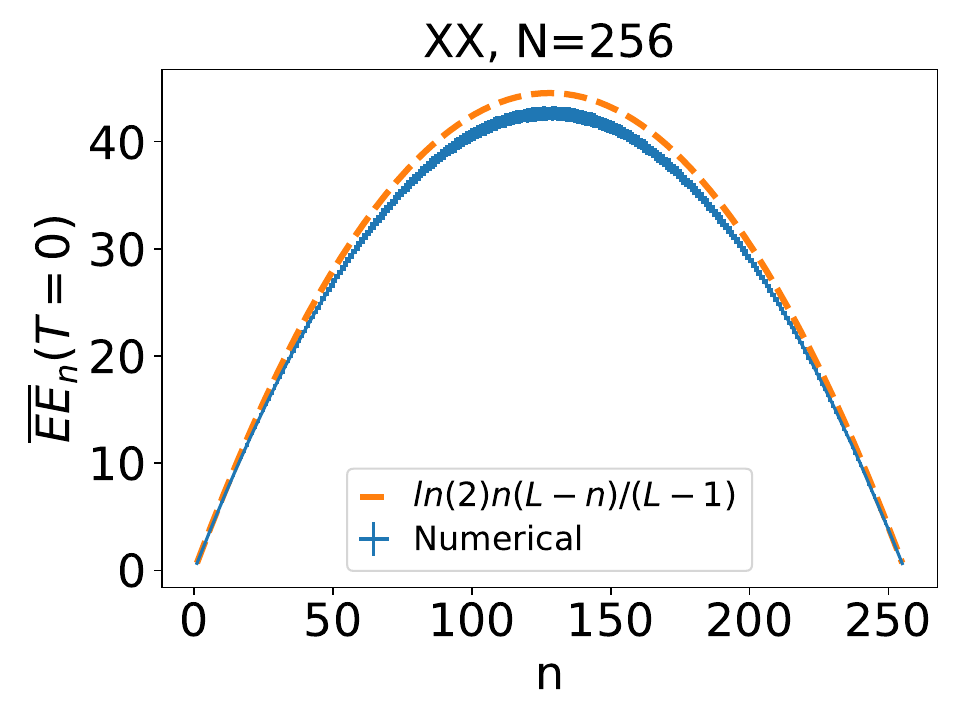}}
    \def\little{\includegraphics[width=0.09\textwidth]{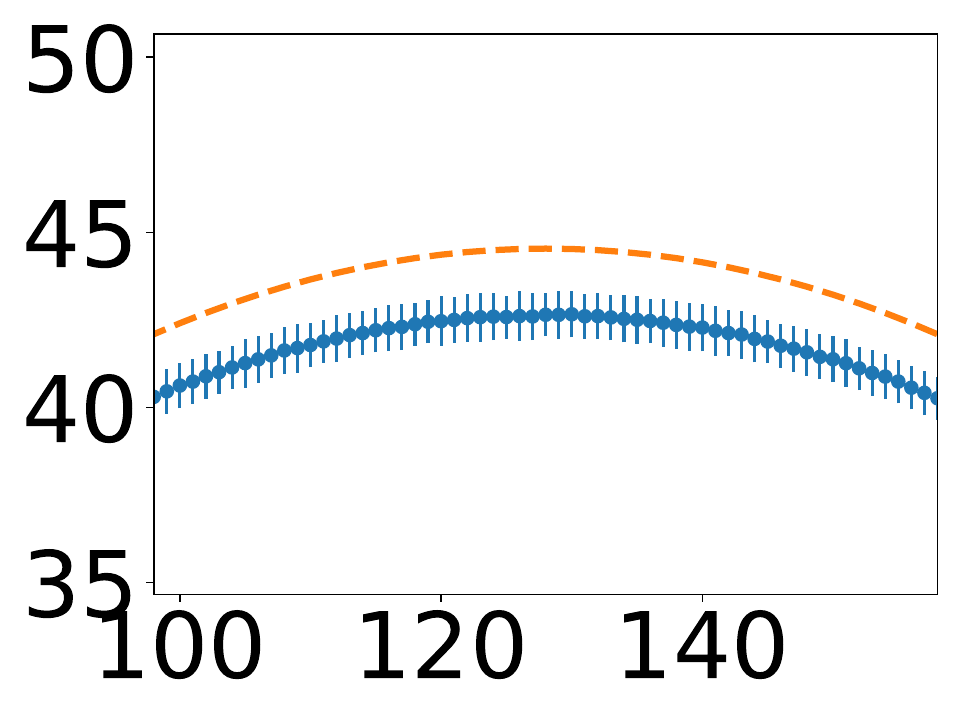}}
    \stackinset{l}{65pt}{t}{40pt}{\little}{\big}
  \end{subfigure}%
  ~%
  \begin{subfigure}{}%
    \def\big{\includegraphics[width=0.31\textwidth]{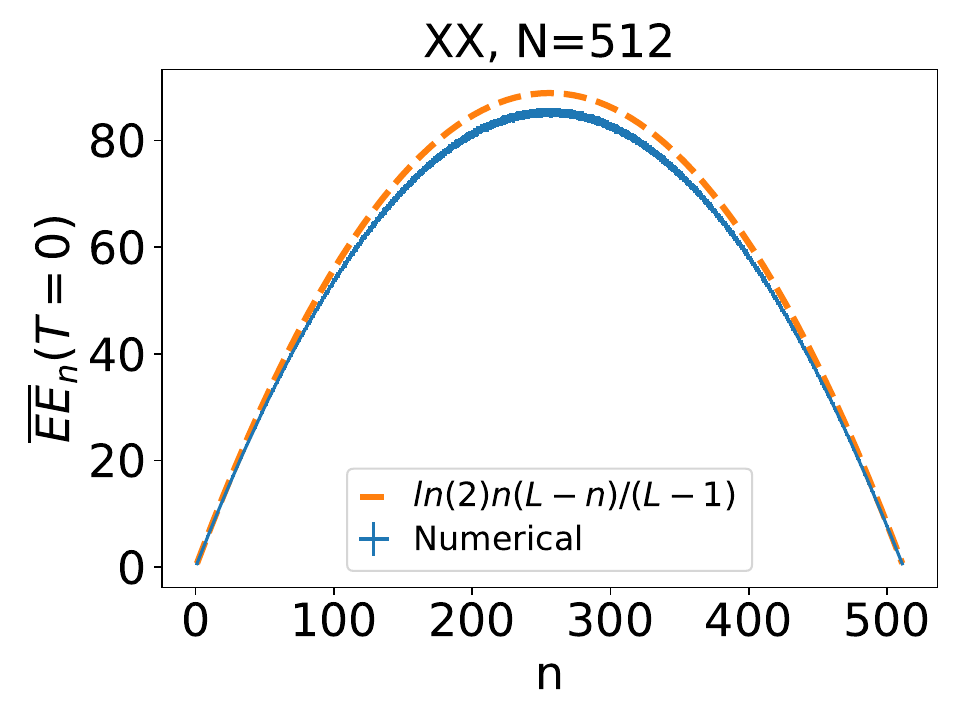}}
    \def\little{\includegraphics[width=0.09\textwidth]{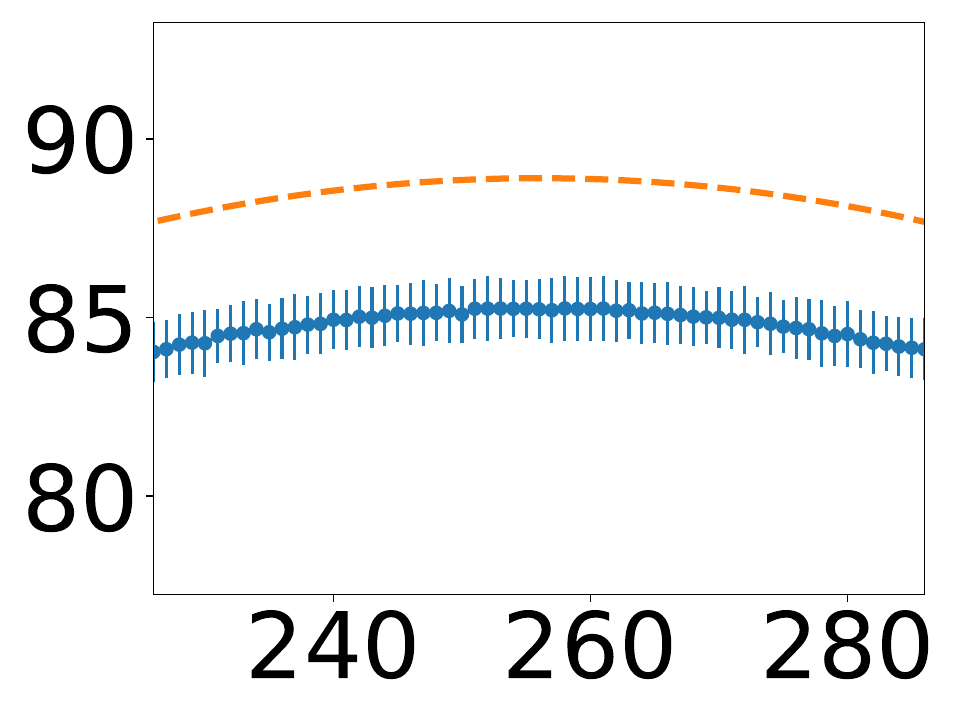}}
    \stackinset{l}{65pt}{t}{40pt}{\little}{\big}
  \end{subfigure}
  ~%
  \begin{subfigure}{}%
    \includegraphics[width=0.31\textwidth]{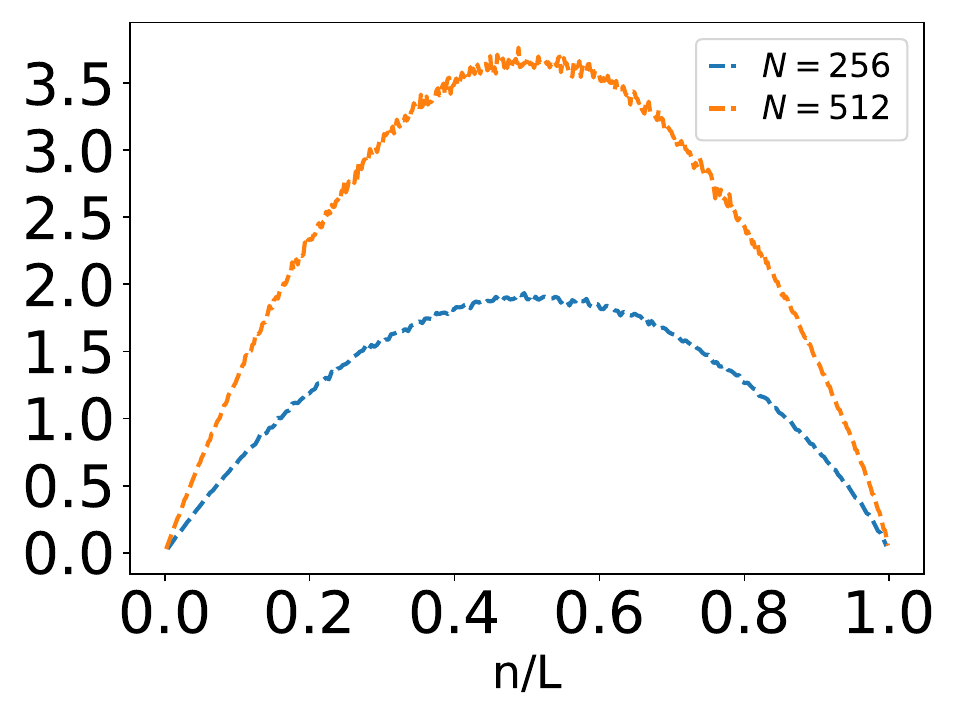}
  \end{subfigure}
  \caption{Plot of $\overline{\text{EE}}$ versus system size $n$ at
    $T=0$. In the left panel, we set $N=256$ and in the middle panel,
    we set $N=512$. Plots are symmetric about $n=L/2$ (standard
    deviations are included). In the right panel, we plot the
    difference between the numerical and analytical results of
    Eq. (\ref{EET0}) versus $n/L$.  The numerical results fairly
    matches the analytical results for subsystem sizes $n$ close to $1$
    and $L$ and it deviates from analytical results about $n \sim L/2$
    (See the inset plots. The deviation is less than $4\%$). For each
    data point, we take disorder average over $\sim 10^{2}$ samples,
    subsystem average over $\sim 10^{2}$ samples, and thus we take the
    average over $\sim 10^{4}$ samples in total.}
  \label{fig:RP_XX_subsys}
\end{figure*}

First, we check the symmetric property of
$\overline{\text{EE}}_n=\overline{\text{EE}}_{L-n}$, meaning that the
$\overline{\text{EE}}_n$ has to be symmetric about $n=L/2$. In
addition, we compare the $\overline{\text{EE}}$ calculated numerically
with Eq.  (\ref{EET0}). As we can see in Fig. \ref{fig:RP_XX_subsys},
the plot of $\overline{\text{EE}}$ numerically obtained is symmetric
about $L/2$, and also it fairly matches with Eq. (\ref{EET0}).

\begin{figure} \centering
  \begin{subfigure}{}%
    \includegraphics[width=0.23\textwidth]{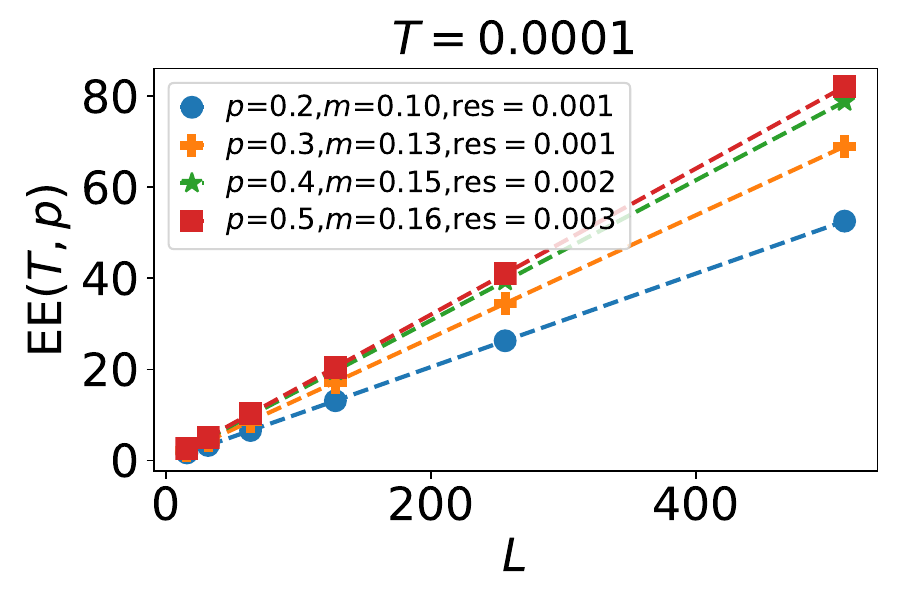}
  \end{subfigure}
  \begin{subfigure}{}%
    \includegraphics[width=0.23\textwidth]{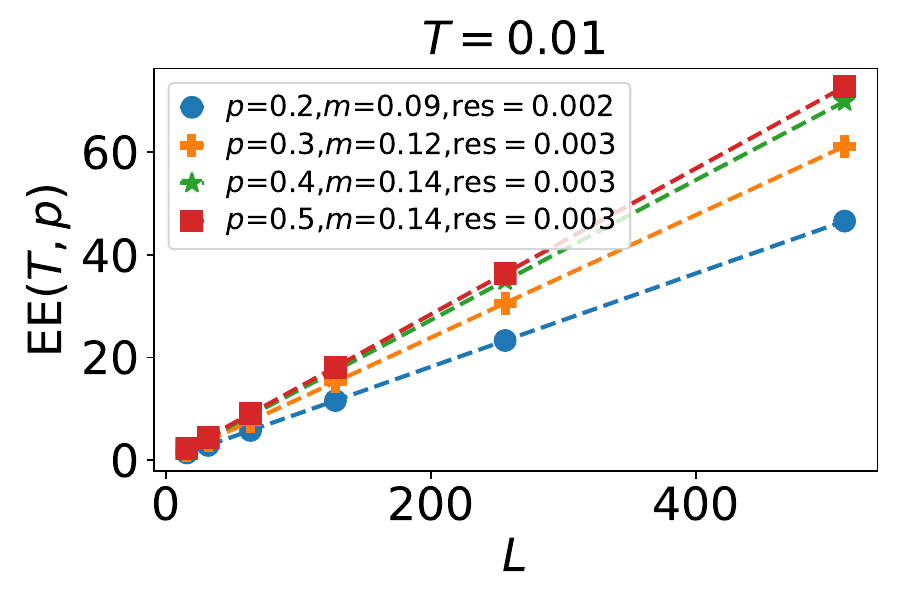}
  \end{subfigure}
  \begin{subfigure}{}%
    \includegraphics[width=0.23\textwidth]{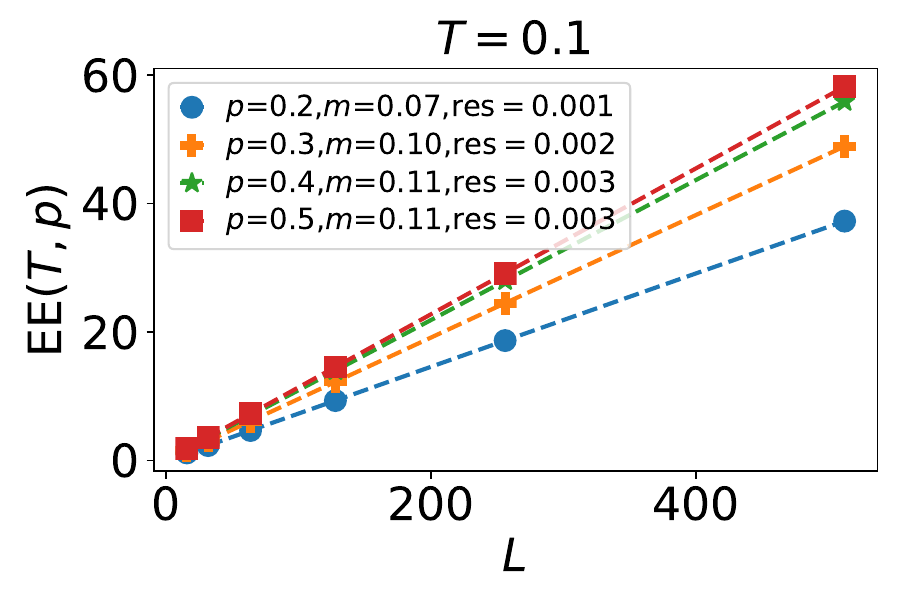}
  \end{subfigure}
  \begin{subfigure}{}%
    \includegraphics[width=0.23\textwidth]{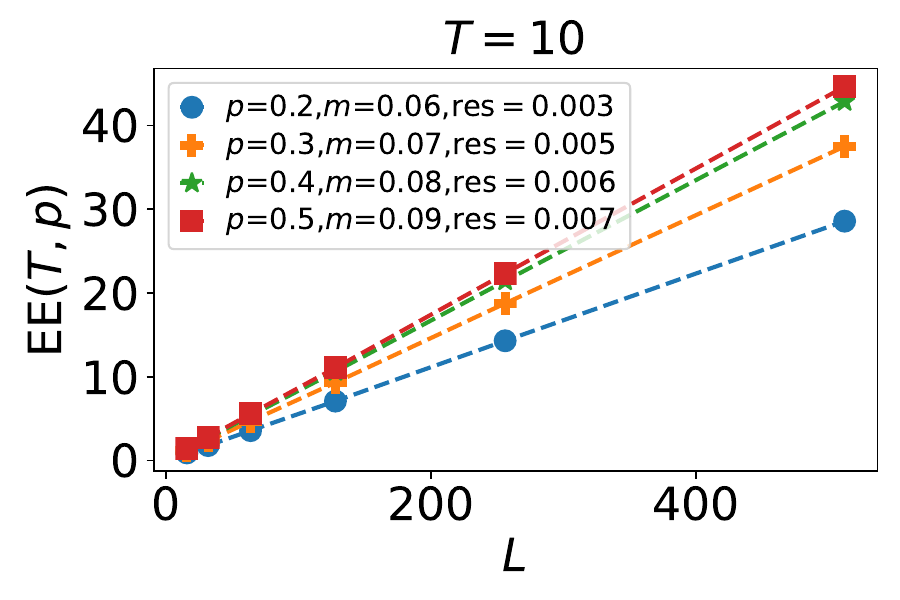}
  \end{subfigure}%
  \caption{Plot of EE$(T,p)$ versus system size $L$ at fixed
    temperatures $T$ for some selected values of uniform probabilities
    $p$. For each probability $p$, the slope of the fitted straight
    line with the obtained numerical results of EE is denoted as
    $m$. Also, the sum of squared residuals of the least squares fits
    is denoted as `res'. Small values for res, denote the fact that
    the straight lines are a good fit for the numerical data. For each
    data point, we take disorder average over $\sim 10^{2}$ samples,
    subsystem average over $\sim 10^{2}$ samples, ensemble average
    over $\sim 10$ samples, and thus we take the average over
    $\sim 10^{5}$ samples in total. \label{fig:EE_L}}
\end{figure}

\begin{figure} \centering
  \begin{subfigure}{}%
    \includegraphics[width=0.23\textwidth]{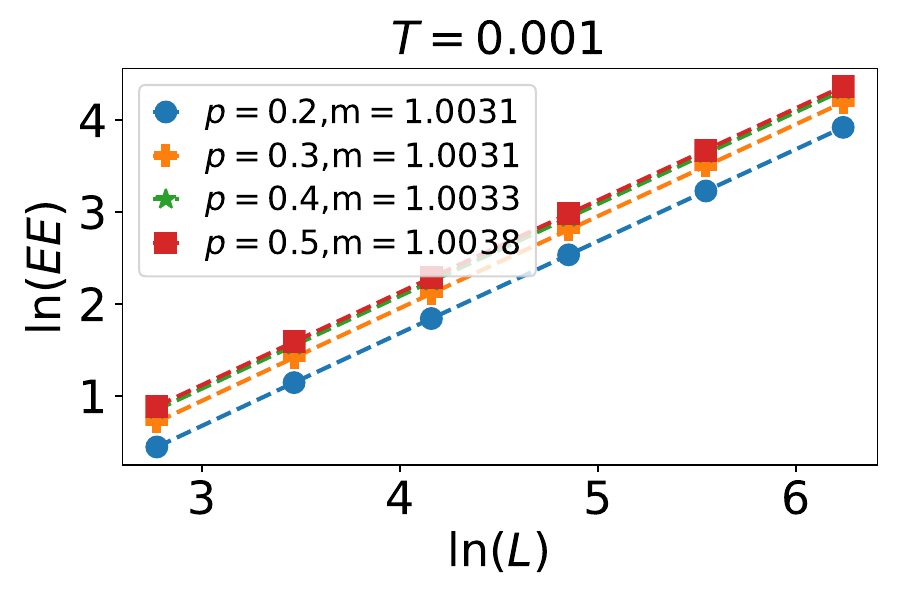}
  \end{subfigure}
  \begin{subfigure}{}%
    \includegraphics[width=0.23\textwidth]{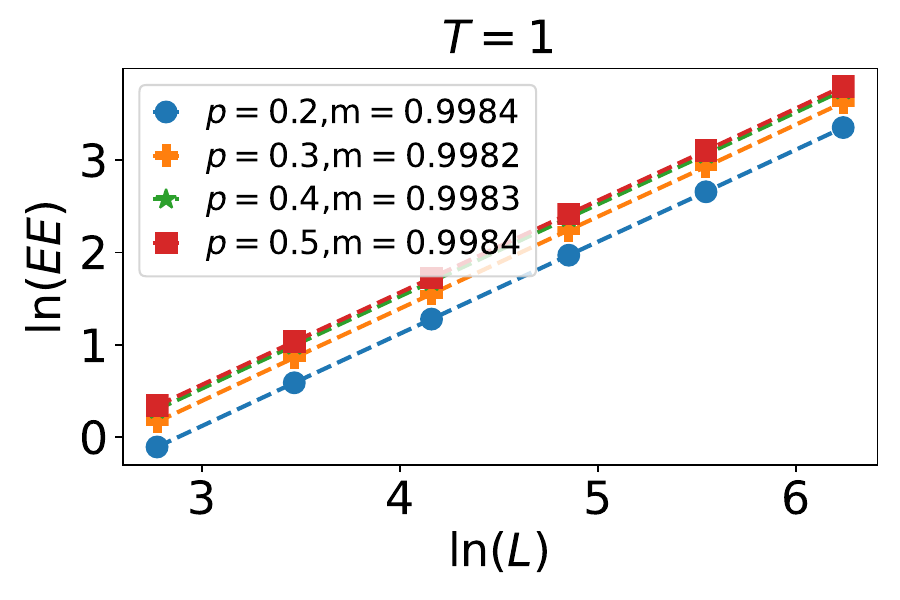}
  \end{subfigure}
  \caption{Plot of the $\ln(\text{EE})$ versus $\ln(L)$ at two fixed
    temperature $T$ for some values of probability $p$. The slope of
    the straight fitted line is denoted as $m$. For each data point,
    we take disorder average over $\sim 10^{2}$ samples, subsystem
    average over $\sim 10^{2}$ samples, ensemble average over
    $\sim 10$ samples and thus we take the average over $\sim 10^{5}$
    samples in total.}
  \label{fig:logEE_logL}
\end{figure}

Next, we check the $L$ dependence of the EE$(T,p)$. We plot the
numerical data of EE versus system size $L$ and the fitted straight
line. This comparison is plotted in Fig. \ref{fig:EE_L} for different
temperatures and probabilities. The sum of squared residuals of the
least squares fits are also denoted; since they have very small
values, we can conclude that the straight-fitted lines are fitted the
numerical data very well. As a double check, we also fit the logarithm
of the EE versus the logarithm of the system size with a straight
line, and we find that the slope is very close to $1$ (see
Fig. \ref{fig:logEE_logL}). Thus, we conclude that $EE \propto L $,
i.e., the EE has power-law behavior for system size $L$.

\begin{figure*}
  \centering
  \begin{subfigure}{}%
    \includegraphics[width=0.32\textwidth]{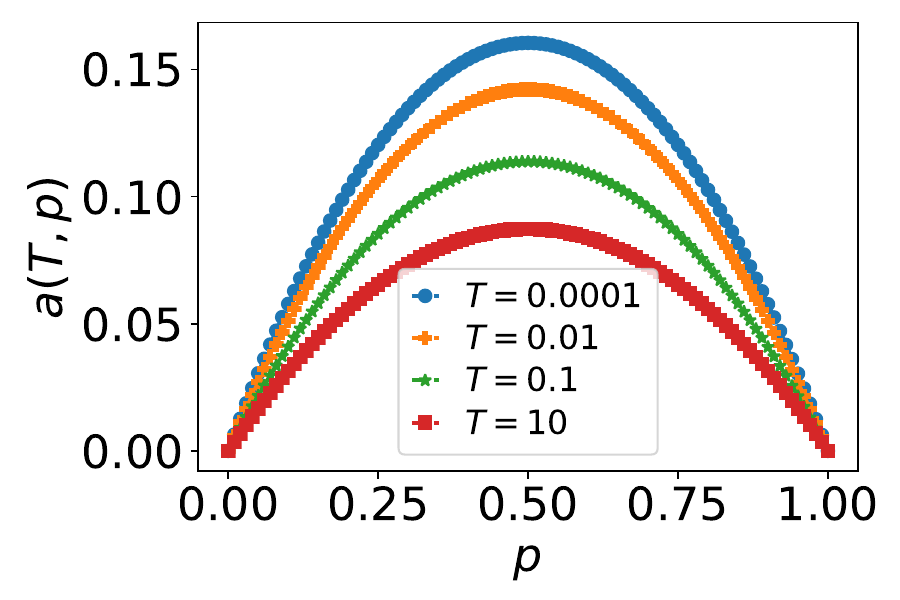}
  \end{subfigure}
  \begin{subfigure}{}%
    \includegraphics[width=0.32\textwidth]{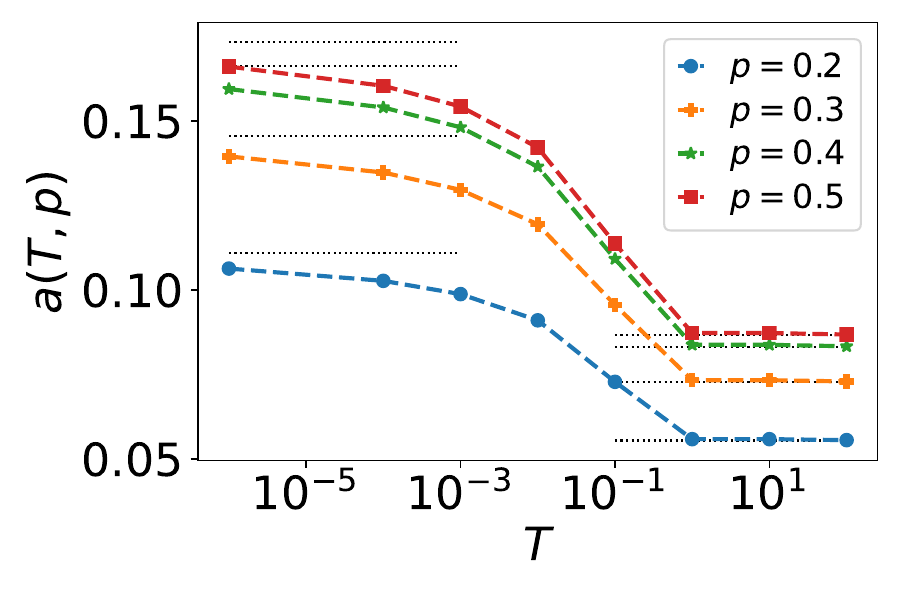}
  \end{subfigure}
  \begin{subfigure}{}%
    \includegraphics[width=0.32\textwidth]{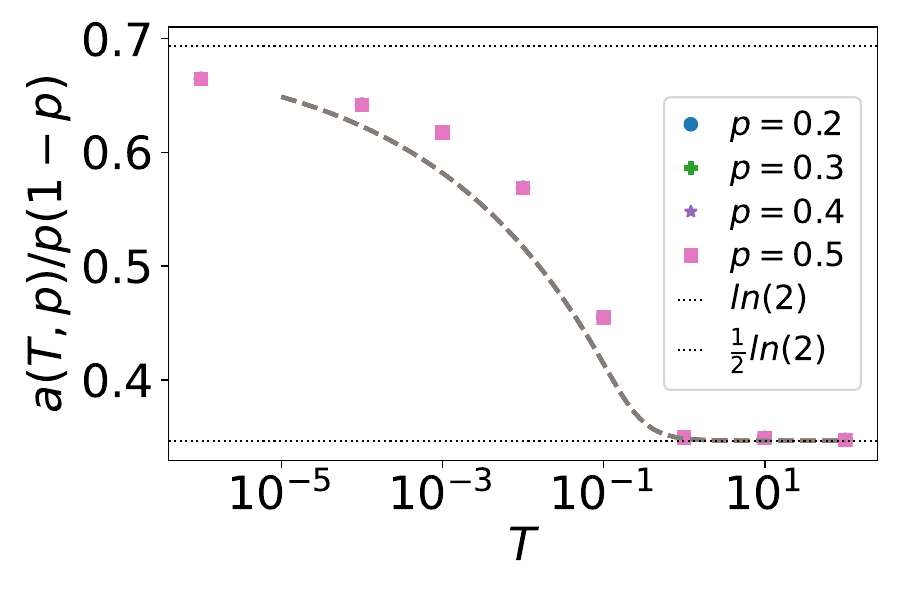}
  \end{subfigure}
  \caption{The EE versus system size $L$ is fitted with the straight
    line $aL+c$, and the results of the behavior of $a$ as a function
    of probability (left panel) and temperature (middle panel) are
    plotted. In the middle panel, the horizontal lines are
    $\ln(2)p(1-p)$ and $\frac{1}{2}\ln(2)p(1-p)$ for each $p$. In the
    right panel, the numerically obtained $\frac{a}{p(1-p)}$ are
    plotted for different probabilities, $p$. We can see that they
    coincide.  In addition, The result of the analytically obtained
    $\frac{a}{p(1-p)}$ based on Eq. (\ref{EEpt}), is plotted with a
    dashed line. These two numerical and analytical results nearly
    match. Two horizontal lines of $\ln(2)$ and $\frac{\ln(2)}{2}$ are
    also plotted. For each data point, we take disorder average over
    $\sim 10^{2}$ samples, subsystem average over $\sim 10^{2}$
    samples, ensemble average over $\sim 10$ samples and thus we take
    the average over $\sim 10^{5}$ samples in total.}
  \label{fig:a}
\end{figure*}

\begin{figure}
  \centering
  \begin{subfigure}{}%
    \includegraphics[width=0.23\textwidth]{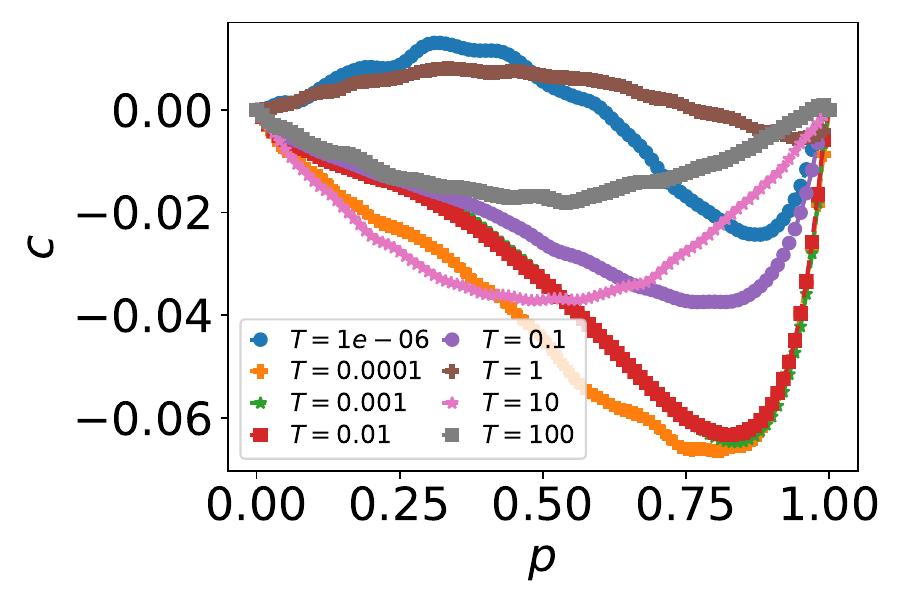}
  \end{subfigure}
  \begin{subfigure}{}%
    \includegraphics[width=0.23\textwidth]{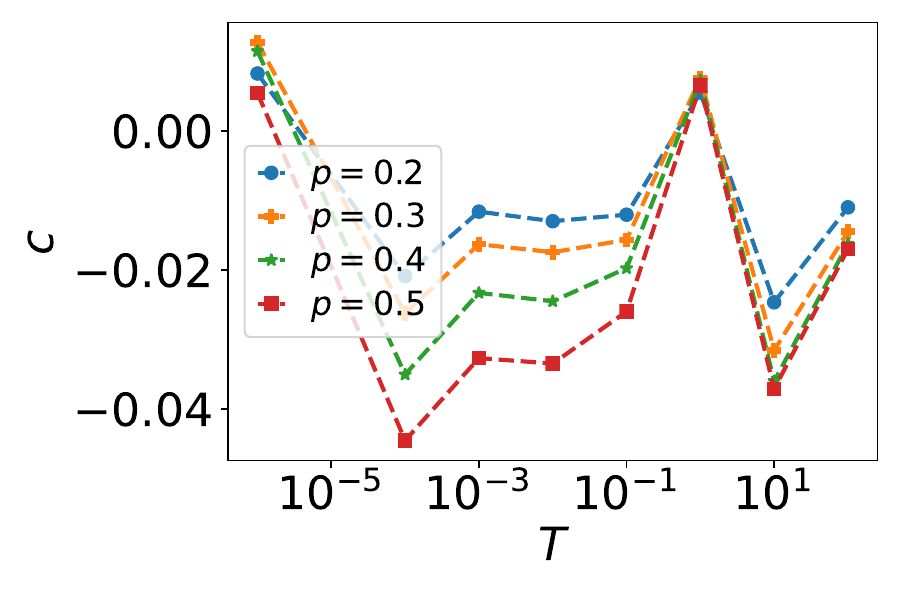}
  \end{subfigure}
  \caption{The EE versus system size $L$ is fitted with the straight
    line $aL+c$, and the results of the y-intercept behavior $c$ as a
    function of the probability (left panel) and also temperature
    (middle panel) are plotted. The values of the numerically obtained
    $c$ are smaller than $10^{-2}$ which is much smaller than the
    value for the EE. For each data point, we take disorder average
    over $\sim 10^{2}$ samples, subsystem average over $\sim 10^{2}$
    samples, ensemble average over $\sim 10$ samples and thus we take
    the average over $\sim 10^{5}$ samples in total.}
  \label{fig:c}
\end{figure}

Now that we know the system size dependence is a power-law with power
$1$, we write EE$=a(T,p) L+c$ (we add the y-intercept $c$ to be
determined with numerical calculations) and study the behavior of
coefficient $a$ and $c$ as a function of temperature $T$ and
probability $p$. One way to do this, is to fit the data of EE versus
$L$ with a straight line and obtain the slope and the y-intercept of
the fitted line numerically. The results of these calculations are
plotted in Figs. \ref{fig:a} and \ref{fig:c}. From the behavior of $a$
versus probability, we can see that it is symmetric about $p=1/2$,
consistent with the analytical result of Eq. (\ref{EEpt}) (see left
panel of Fig. \ref{fig:a}).

In addition, from the behavior of $a$ versus temperature, we can see
that it approaches constant values at low and high temperatures. These
constants are consistent with the analytical values of $a$ in
Eq. (\ref{EEpt1}), which are $\ln(2) \times p(1-p)$ in low
temperatures, and $\frac{1}{2} \ln(2)\times p(1-p)$ for high
temperatures (Since $\langle P_s + P_{t_{\uparrow\downarrow}} \rangle$
goes to $1$ in the low $T$ limit and it goes to $\frac{1}{2}$ in the
high $T$ limit. See the middle panel of Fig. \ref{fig:a}).

In the right panel of Fig. \ref{fig:a}, we do the following. First,
for the numerically obtained values of $a$, we plot $\frac{a}{p(1-p)}$
for some selected values of $p$. As we can see, they all coincide with
each other. I.e., the only probability dependence is in the form of
$p(1-p)$. In addition, we can see that the numerically obtained value
of $a$ goes to $\ln(2)$ in the low $T$ limit, and it goes to
$\frac{1}{2} \ln(2)$ in the high $T$ limit, which are consistent with
the analytical result of the
$\frac{a}{p(1-p)} = \ln(2) \langle P_s + P_{t_{\uparrow\downarrow}}
\rangle$. For an arbitrary temperature $T$, numerical data and the
analytical predictions of $a$ are in a fair agreement. The difference
between the numerically and analytically obtained values of $a$ stems
from the fact that we use an approximate RSRG method to calculate the
EE. We also note that, in a numerical calculation, it is always
possible to benefit from larger system sizes to avoid finite-size
scaling. Finally, we can see that the numerically obtained values of
$c$ plotted in Fig. \ref{fig:c}, are very small compared to the EE
values; they are thus negligible.

\section{conclusion and outlook}\label{sec:conclusion}
\begin{figure}
  \begin{subfigure}{}%
    \includegraphics[width=0.23\textwidth]{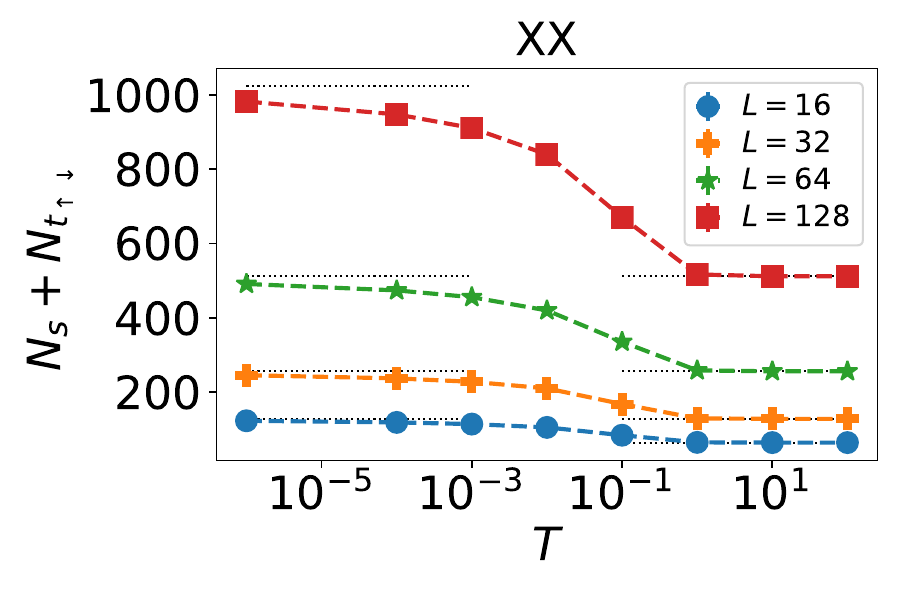}
  \end{subfigure}%
  \begin{subfigure}{}%
    \includegraphics[width=0.23\textwidth]{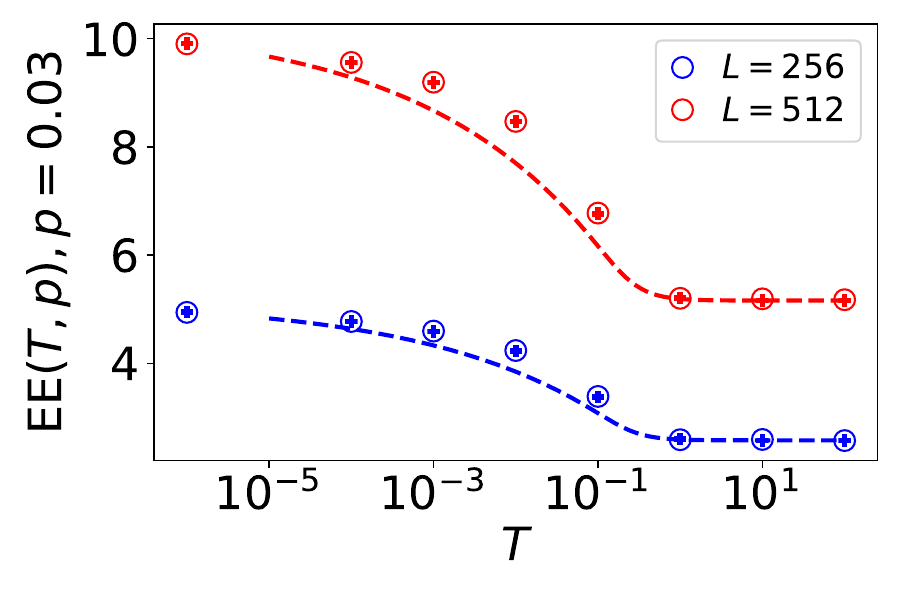}
  \end{subfigure}%
  \caption{Left panel: Number of singlets and
    triplet$_{\uparrow\downarrow}$s forming all over the entire system
    versus temperature $T$. Horizontal lines are $L/2$ and
    $L/8$. Right panel: three different data are plotted: the
    numerically obtained data of the EE denoted by an unfilled
    circles,
    $\ln(2) \langle P_s + P_{t_{\uparrow\downarrow}} \rangle p(1-p) L$
    with a dashed line, and
    $2\ln(2) p(1-p) (N_s + N_{t_{\uparrow\downarrow}})$ with a plus
    sign. For each data point, we take disorder average over
    $\sim 10^{2}$ samples, subsystem average over $\sim 10^{2}$
    samples, ensemble average over $\sim 10$ samples and thus we take
    the average over $\sim 10^{5}$ samples in total. \label{fig:NST}}
\end{figure}

The usual way of measuring the entanglement properties of a system is
to bipartite it into two subsystems, and then obtain the
\emph{non-local} entanglement properties by calculating the EE. What
we did in this paper is distinct: each site has a chance to be part of
the subsystem, subsystem size is also varying, and in addition, for
each subsystem size, we do all different ways of partitioning, and
then we take the average over the calculated EE's. In such wise, we
measure long-range as well as short-range correlations in the
system. In other words, when we calculate the EE for a bi-partitioned
system, we measure how much subsystem is entangled with the
environment, in which long-range correlations and also
near-to-the-boundary short-range correlations take part. On the other
hand, we are measuring both the short-range and long-range
correlations in the entire system when we randomly partition the
system in all possible ways,

Considering the XX spin chain, if we cut the system in the middle and
then calculate the EE, we are counting the number of entangled bonds
(singlet and triplet$_{\uparrow\downarrow}$) that cross the middle of
the system, and thus we are counting the long bonds or those short
bonds that are close to the boundary. But, since in random
partitioning the partitions are random and they can be disconnected as
well, we are counting the entangled bonds, both with short and long
lengths. In averaging over all such partitioning, we thus count the
number of singlets and triplet$_{\uparrow\downarrow}$s \emph{all over
  the entire system}. This is, of course, in agreement with the
analytical point of view that
$N_s + N_{t_{\uparrow\downarrow}} = \langle P_s +
P_{t_{\uparrow\downarrow}} \rangle \times \frac{L}{2}$, and thus we
can rewrite the EE$(T,p)$ of Eq. (\ref{EEpt}) as the following:
\begin{equation}\label{EETpNst}
  \text{EE}(T, p) =2\ln(2) p(1-p)  (N_s + N_{t_{\uparrow\downarrow}})
\end{equation}
In Fig. \ref{fig:NST}, we plot the number of singlets and
triplet$_{\uparrow\downarrow}$s forming in the entire system as a
function of the temperature. In the low $T$ limit, all bonds are only
singlet and triplet$_{\uparrow\downarrow}$, and the sum goes to
$L/2$. On the other hand, in the high-temperature limit, the singlet
and the three triplets have the same probability, so the sum goes to
$L/8$. As a numerical check, we also plot and compare
Eq. (\ref{EETpNst}) with the numerically obtained data for the EE. We
see full agreement.

The XX spin chain, which we employed in this paper, gives us a
schematic representation of the bonds forming in the system. However,
this picture is not always available.  So in general, to see both
short- and long- range-correlations, we can use the random
partitioning method. In particular, the behavior of the EE in a
Hamiltonian with local and non-local interactions would be
interesting. The random partitioning can also be used in
characterizations of the phase transition. For example, in the
Anderson delocalized-localized phase
transition\cite{https://doi.org/10.48550/arxiv.2208.11409}. Besides,
we only considered the uniform probability distribution, i.e., each
site has the same chance to belong to the subsystem. Considering
non-uniform probabilities could also be helpful and give us more
physical insights.

\section*{acknowledgments}
The author thanks Dr. Jahanfar Abouei for valuable discussions.  The
author gratefully acknowledges the high performance computing center
of the university of Mazandaran for providing computing resources and
time. This work is based upon research funded by Iran National Science Foundation (INSF) under project No. 4000258.


\begin{thebibliography}{10}
	\expandafter\ifx\csname url\endcsname\relax
	\def\url#1{\texttt{#1}}\fi
	\expandafter\ifx\csname urlprefix\endcsname\relax\def\urlprefix{URL }\fi
	\expandafter\ifx\csname href\endcsname\relax
	\def\href#1#2{#2} \def\path#1{#1}\fi

	\bibitem{RevModPhys.81.865}
	R.~Horodecki, P.~Horodecki, M.~Horodecki, K.~Horodecki,
	\href{https://link.aps.org/doi/10.1103/RevModPhys.81.865}{Quantum
		entanglement}, Rev. Mod. Phys. 81 (2009) 865--942.
	\newblock \href {https://doi.org/10.1103/RevModPhys.81.865}
	{\path{doi:10.1103/RevModPhys.81.865}}.
	\newline\urlprefix\url{https://link.aps.org/doi/10.1103/RevModPhys.81.865}

	\bibitem{LAFLORENCIE20161}
	N.~Laflorencie,
	\href{http://www.sciencedirect.com/science/article/pii/S0370157316301582}{Quantum
		entanglement in condensed matter systems}, Physics Reports 646 (2016) 1 --
	59.
	\newblock \href {https://doi.org/https://doi.org/10.1016/j.physrep.2016.06.008}
	{\path{doi:https://doi.org/10.1016/j.physrep.2016.06.008}}.
	\newline\urlprefix\url{http://www.sciencedirect.com/science/article/pii/S0370157316301582}

	\bibitem{Kauffman_2002}
	L.~H. Kauffman, S.~J. Lomonaco,
	\href{https://doi.org/10.1088/1367-2630/4/1/373}{Quantum entanglement and
		topological entanglement}, New Journal of Physics 4 (2002) 73--73.
	\newblock \href {https://doi.org/10.1088/1367-2630/4/1/373}
	{\path{doi:10.1088/1367-2630/4/1/373}}.
	\newline\urlprefix\url{https://doi.org/10.1088/1367-2630/4/1/373}

	\bibitem{PhysRevX.7.021021}
	D.-L. Deng, X.~Li, S.~Das~Sarma,
	\href{https://link.aps.org/doi/10.1103/PhysRevX.7.021021}{Quantum
		entanglement in neural network states}, Phys. Rev. X 7 (2017) 021021.
	\newblock \href {https://doi.org/10.1103/PhysRevX.7.021021}
	{\path{doi:10.1103/PhysRevX.7.021021}}.
	\newline\urlprefix\url{https://link.aps.org/doi/10.1103/PhysRevX.7.021021}

	\bibitem{PhysRevLett.96.110404}
	A.~Kitaev, J.~Preskill,
	\href{https://link.aps.org/doi/10.1103/PhysRevLett.96.110404}{Topological
		entanglement entropy}, Phys. Rev. Lett. 96 (2006) 110404.
	\newblock \href {https://doi.org/10.1103/PhysRevLett.96.110404}
	{\path{doi:10.1103/PhysRevLett.96.110404}}.
	\newline\urlprefix\url{https://link.aps.org/doi/10.1103/PhysRevLett.96.110404}

	\bibitem{Calabrese_2004}
	P.~Calabrese, J.~Cardy,
	\href{https://doi.org/10.1088/1742-5468/2004/06/p06002}{Entanglement entropy
		and quantum field theory}, Journal of Statistical Mechanics: Theory and
	Experiment 2004~(06) (2004) P06002.
	\newblock \href {https://doi.org/10.1088/1742-5468/2004/06/p06002}
	{\path{doi:10.1088/1742-5468/2004/06/p06002}}.
	\newline\urlprefix\url{https://doi.org/10.1088/1742-5468/2004/06/p06002}

	\bibitem{PhysRev.47.777}
	A.~Einstein, B.~Podolsky, N.~Rosen,
	\href{https://link.aps.org/doi/10.1103/PhysRev.47.777}{Can quantum-mechanical
		description of physical reality be considered complete?}, Phys. Rev. 47
	(1935) 777--780.
	\newblock \href {https://doi.org/10.1103/PhysRev.47.777}
	{\path{doi:10.1103/PhysRev.47.777}}.
	\newline\urlprefix\url{https://link.aps.org/doi/10.1103/PhysRev.47.777}

	\bibitem{schrodinger_1935}
	E.~Schr\"{o}dinger, Discussion of probability relations between separated
	systems, Mathematical Proceedings of the Cambridge Philosophical Society
	31~(4) (1935) 555–563.
	\newblock \href {https://doi.org/10.1017/S0305004100013554}
	{\path{doi:10.1017/S0305004100013554}}.

	\bibitem{PhysRevA.92.042329}
	S.~Szalay,
	\href{https://link.aps.org/doi/10.1103/PhysRevA.92.042329}{Multipartite
		entanglement measures}, Phys. Rev. A 92 (2015) 042329.
	\newblock \href {https://doi.org/10.1103/PhysRevA.92.042329}
	{\path{doi:10.1103/PhysRevA.92.042329}}.
	\newline\urlprefix\url{https://link.aps.org/doi/10.1103/PhysRevA.92.042329}

	\bibitem{PhysRevLett.78.2275}
	V.~Vedral, M.~B. Plenio, M.~A. Rippin, P.~L. Knight,
	\href{https://link.aps.org/doi/10.1103/PhysRevLett.78.2275}{Quantifying
		entanglement}, Phys. Rev. Lett. 78 (1997) 2275--2279.
	\newblock \href {https://doi.org/10.1103/PhysRevLett.78.2275}
	{\path{doi:10.1103/PhysRevLett.78.2275}}.
	\newline\urlprefix\url{https://link.aps.org/doi/10.1103/PhysRevLett.78.2275}

	\bibitem{RevModPhys.73.565}
	J.~M. Raimond, M.~Brune, S.~Haroche,
	\href{https://link.aps.org/doi/10.1103/RevModPhys.73.565}{Manipulating
		quantum entanglement with atoms and photons in a cavity}, Rev. Mod. Phys. 73
	(2001) 565--582.
	\newblock \href {https://doi.org/10.1103/RevModPhys.73.565}
	{\path{doi:10.1103/RevModPhys.73.565}}.
	\newline\urlprefix\url{https://link.aps.org/doi/10.1103/RevModPhys.73.565}

	\bibitem{PhysRevA.99.062309}
	E.~Cornfeld, E.~Sela, M.~Goldstein,
	\href{https://link.aps.org/doi/10.1103/PhysRevA.99.062309}{Measuring
		fermionic entanglement: Entropy, negativity, and spin structure}, Phys. Rev.
	A 99 (2019) 062309.
	\newblock \href {https://doi.org/10.1103/PhysRevA.99.062309}
	{\path{doi:10.1103/PhysRevA.99.062309}}.
	\newline\urlprefix\url{https://link.aps.org/doi/10.1103/PhysRevA.99.062309}

	\bibitem{Sackett2000}
	C.~A. Sackett, D.~Kielpinski, B.~E. King, C.~Langer, V.~Meyer, C.~J. Myatt,
	M.~Rowe, Q.~A. Turchette, W.~M. Itano, D.~J. Wineland, C.~Monroe,
	\href{http://dx.doi.org/10.1038/35005011}{Experimental entanglement of four
		particles}, Nature 404 (2000) 256 EP --.
	\newline\urlprefix\url{http://dx.doi.org/10.1038/35005011}

	\bibitem{RevModPhys.82.277}
	J.~Eisert, M.~Cramer, M.~B. Plenio,
	\href{https://link.aps.org/doi/10.1103/RevModPhys.82.277}{Colloquium: Area
		laws for the entanglement entropy}, Rev. Mod. Phys. 82 (2010) 277--306.
	\newblock \href {https://doi.org/10.1103/RevModPhys.82.277}
	{\path{doi:10.1103/RevModPhys.82.277}}.
	\newline\urlprefix\url{https://link.aps.org/doi/10.1103/RevModPhys.82.277}

	\bibitem{PhysRevLett.96.010404}
	M.~M. Wolf,
	\href{https://link.aps.org/doi/10.1103/PhysRevLett.96.010404}{Violation of
		the entropic area law for fermions}, Phys. Rev. Lett. 96 (2006) 010404.
	\newblock \href {https://doi.org/10.1103/PhysRevLett.96.010404}
	{\path{doi:10.1103/PhysRevLett.96.010404}}.
	\newline\urlprefix\url{https://link.aps.org/doi/10.1103/PhysRevLett.96.010404}

	\bibitem{Vitagliano_2010}
	G.~Vitagliano, A.~Riera, J.~I. Latorre,
	\href{https://doi.org/10.1088/1367-2630/12/11/113049}{Volume-law scaling for
		the entanglement entropy in spin-1/2 chains}, New Journal of Physics 12~(11)
	(2010) 113049.
	\newblock \href {https://doi.org/10.1088/1367-2630/12/11/113049}
	{\path{doi:10.1088/1367-2630/12/11/113049}}.
	\newline\urlprefix\url{https://doi.org/10.1088/1367-2630/12/11/113049}

	\bibitem{PhysRevB.89.115104}
	M.~Pouranvari, K.~Yang,
	\href{https://link.aps.org/doi/10.1103/PhysRevB.89.115104}{Maximally
		entangled mode, metal-insulator transition, and violation of entanglement
		area law in noninteracting fermion ground states}, Phys. Rev. B 89 (2014)
	115104.
	\newblock \href {https://doi.org/10.1103/PhysRevB.89.115104}
	{\path{doi:10.1103/PhysRevB.89.115104}}.
	\newline\urlprefix\url{https://link.aps.org/doi/10.1103/PhysRevB.89.115104}

	\bibitem{PhysRevLett.109.267203}
	T.~Koffel, M.~Lewenstein, L.~Tagliacozzo,
	\href{https://link.aps.org/doi/10.1103/PhysRevLett.109.267203}{Entanglement
		entropy for the long-range ising chain in a transverse field}, Phys. Rev.
	Lett. 109 (2012) 267203.
	\newblock \href {https://doi.org/10.1103/PhysRevLett.109.267203}
	{\path{doi:10.1103/PhysRevLett.109.267203}}.
	\newline\urlprefix\url{https://link.aps.org/doi/10.1103/PhysRevLett.109.267203}

	\bibitem{PhysRevLett.105.050502}
	B.~Swingle,
	\href{https://link.aps.org/doi/10.1103/PhysRevLett.105.050502}{Entanglement
		entropy and the fermi surface}, Phys. Rev. Lett. 105 (2010) 050502.
	\newblock \href {https://doi.org/10.1103/PhysRevLett.105.050502}
	{\path{doi:10.1103/PhysRevLett.105.050502}}.
	\newline\urlprefix\url{https://link.aps.org/doi/10.1103/PhysRevLett.105.050502}

	\bibitem{Wong2013}
	G.~Wong, I.~Klich, L.~A.~P. Zayas, D.~Vaman,
	\href{https://doi.org/10.1007/JHEP12(2013)020}{Entanglement temperature and
		entanglement entropy of excited states}, Journal of High Energy Physics
	2013~(12) (2013) 20.
	\newblock \href {https://doi.org/10.1007/JHEP12(2013)020}
	{\path{doi:10.1007/JHEP12(2013)020}}.
	\newline\urlprefix\url{https://doi.org/10.1007/JHEP12(2013)020}

	\bibitem{Alba_2009}
	V.~Alba, M.~Fagotti, P.~Calabrese,
	\href{https://doi.org/10.1088%2F1742-5468%2F2009%2F10%2Fp10020}{Entanglement
		entropy of excited states}, Journal of Statistical Mechanics: Theory and
	Experiment 2009~(10) (2009) P10020.
	\newblock \href {https://doi.org/10.1088/1742-5468/2009/10/p10020}
	{\path{doi:10.1088/1742-5468/2009/10/p10020}}.
	\newline\urlprefix\url{https://doi.org/10.1088%2F1742-5468%2F2009%2F10%2Fp10020}

	\bibitem{PhysRevB.90.220202}
	Y.~Huang, J.~E. Moore,
	\href{https://link.aps.org/doi/10.1103/PhysRevB.90.220202}{Excited-state
		entanglement and thermal mutual information in random spin chains}, Phys.
	Rev. B 90 (2014) 220202.
	\newblock \href {https://doi.org/10.1103/PhysRevB.90.220202}
	{\path{doi:10.1103/PhysRevB.90.220202}}.
	\newline\urlprefix\url{https://link.aps.org/doi/10.1103/PhysRevB.90.220202}

	\bibitem{PhysRevLett.110.091602}
	J.~Bhattacharya, M.~Nozaki, T.~Takayanagi, T.~Ugajin,
	\href{https://link.aps.org/doi/10.1103/PhysRevLett.110.091602}{Thermodynamical
		property of entanglement entropy for excited states}, Phys. Rev. Lett. 110
	(2013) 091602.
	\newblock \href {https://doi.org/10.1103/PhysRevLett.110.091602}
	{\path{doi:10.1103/PhysRevLett.110.091602}}.
	\newline\urlprefix\url{https://link.aps.org/doi/10.1103/PhysRevLett.110.091602}

	\bibitem{PhysRevB.100.165135}
	A.~Jafarizadeh, M.~A. Rajabpour,
	\href{https://link.aps.org/doi/10.1103/PhysRevB.100.165135}{Bipartite
		entanglement entropy of the excited states of free fermions and harmonic
		oscillators}, Phys. Rev. B 100 (2019) 165135.
	\newblock \href {https://doi.org/10.1103/PhysRevB.100.165135}
	{\path{doi:10.1103/PhysRevB.100.165135}}.
	\newline\urlprefix\url{https://link.aps.org/doi/10.1103/PhysRevB.100.165135}

	\bibitem{Rodriguez-Laguna_2017}
	J.~Rodríguez-Laguna, J.~Dubail, G.~Ramírez, P.~Calabrese, G.~Sierra,
	\href{https://dx.doi.org/10.1088/1751-8121/aa6268}{More on the rainbow chain:
		entanglement, space-time geometry and thermal states}, Journal of Physics A:
	Mathematical and Theoretical 50~(16) (2017) 164001.
	\newblock \href {https://doi.org/10.1088/1751-8121/aa6268}
	{\path{doi:10.1088/1751-8121/aa6268}}.
	\newline\urlprefix\url{https://dx.doi.org/10.1088/1751-8121/aa6268}

	\bibitem{PhysRevB.101.205121}
	N.~Samos S\'aenz~de Buruaga, S.~N. Santalla, J.~Rodr\'{\i}guez-Laguna,
	G.~Sierra,
	\href{https://link.aps.org/doi/10.1103/PhysRevB.101.205121}{Piercing the
		rainbow state: Entanglement on an inhomogeneous spin chain with a defect},
	Phys. Rev. B 101 (2020) 205121.
	\newblock \href {https://doi.org/10.1103/PhysRevB.101.205121}
	{\path{doi:10.1103/PhysRevB.101.205121}}.
	\newline\urlprefix\url{https://link.aps.org/doi/10.1103/PhysRevB.101.205121}

	\bibitem{Ramirez_2014}
	G.~Ramírez, J.~Rodríguez-Laguna, G.~Sierra,
	\href{https://dx.doi.org/10.1088/1742-5468/2014/07/P07003}{Entanglement in
		low-energy states of the random-hopping model}, Journal of Statistical
	Mechanics: Theory and Experiment 2014~(7) (2014) P07003.
	\newblock \href {https://doi.org/10.1088/1742-5468/2014/07/P07003}
	{\path{doi:10.1088/1742-5468/2014/07/P07003}}.
	\newline\urlprefix\url{https://dx.doi.org/10.1088/1742-5468/2014/07/P07003}
	
	
	\bibitem{ahmadi2021frustrated}
	N.~Ahmadi, J.~Abouie, R.~Haghshenas, Frustrated mixed-spin ladders: An
	intermediate phase between rung-singlet and haldane phases (2021).
	\newblock \href {http://arxiv.org/abs/2106.07940} {\path{arXiv:2106.07940}}.

	\bibitem{Moradi_2016}
	Z.~Moradi, J.~Abouie,
	\href{https://doi.org/10.1088%2F1742-5468%2F2016%2F11%2F113101}{Entanglement
		spectrum of fermionic bilayer honeycomb lattice: Hofstadter butterfly},
	Journal of Statistical Mechanics: Theory and Experiment 2016~(11) (2016)
	113101.
	\newblock \href {https://doi.org/10.1088/1742-5468/2016/11/113101}
	{\path{doi:10.1088/1742-5468/2016/11/113101}}.
	\newline\urlprefix\url{https://doi.org/10.1088%2F1742-5468%2F2016%2F11%2F113101}

	\bibitem{PhysRevB.101.195117}
	N.~Ahmadi, J.~Abouie, D.~Baeriswyl,
	\href{https://link.aps.org/doi/10.1103/PhysRevB.101.195117}{Topological and
		nontopological features of generalized su-schrieffer-heeger models}, Phys.
	Rev. B 101 (2020) 195117.
	\newblock \href {https://doi.org/10.1103/PhysRevB.101.195117}
	{\path{doi:10.1103/PhysRevB.101.195117}}.
	\newline\urlprefix\url{https://link.aps.org/doi/10.1103/PhysRevB.101.195117}

	\bibitem{Roosz2020}
	G.~Ro{\'o}sz, I.~A. Kov{\'a}cs, F.~Igl{\'o}i,
	\href{https://doi.org/10.1140/epjb/e2019-100496-y}{Entanglement entropy of
		random partitioning}, The European Physical Journal B 93~(1) (2020) 8.
	\newblock \href {https://doi.org/10.1140/epjb/e2019-100496-y}
	{\path{doi:10.1140/epjb/e2019-100496-y}}.
	\newline\urlprefix\url{https://doi.org/10.1140/epjb/e2019-100496-y}

	\bibitem{PhysRevB.91.220101}
	S.~Vijay, L.~Fu,
	\href{https://link.aps.org/doi/10.1103/PhysRevB.91.220101}{Entanglement
		spectrum of a random partition: Connection with the localization transition},
	Phys. Rev. B 91 (2015) 220101.
	\newblock \href {https://doi.org/10.1103/PhysRevB.91.220101}
	{\path{doi:10.1103/PhysRevB.91.220101}}.
	\newline\urlprefix\url{https://link.aps.org/doi/10.1103/PhysRevB.91.220101}

	\bibitem{PhysRevB.83.045110}
	M.~Fagotti, P.~Calabrese, J.~E. Moore,
	\href{https://link.aps.org/doi/10.1103/PhysRevB.83.045110}{Entanglement
		spectrum of random-singlet quantum critical points}, Phys. Rev. B 83 (2011)
	045110.
	\newblock \href {https://doi.org/10.1103/PhysRevB.83.045110}
	{\path{doi:10.1103/PhysRevB.83.045110}}.
	\newline\urlprefix\url{https://link.aps.org/doi/10.1103/PhysRevB.83.045110}

	\bibitem{Refael_2009}
	G.~Refael, J.~E. Moore,
	\href{https://doi.org/10.1088%2F1751-8113%2F42%2F50%2F504010}{Criticality and
		entanglement in random quantum systems}, Journal of Physics A: Mathematical
	and Theoretical 42~(50) (2009) 504010.
	\newblock \href {https://doi.org/10.1088/1751-8113/42/50/504010}
	{\path{doi:10.1088/1751-8113/42/50/504010}}.
	\newline\urlprefix\url{https://doi.org/10.1088%2F1751-8113%2F42%2F50%2F504010}

	\bibitem{mohdeb2022excitedeigenstate}
	Y.~Mohdeb, J.~Vahedi, S.~Kettemann, Excited-eigenstate entanglement properties
	of xx spin chains with random long-range interactions (2022).
	\newblock \href {http://arxiv.org/abs/2201.10607} {\path{arXiv:2201.10607}}.

	\bibitem{PhysRevB.72.140408}
	N.~Laflorencie,
	\href{https://link.aps.org/doi/10.1103/PhysRevB.72.140408}{Scaling of
		entanglement entropy in the random singlet phase}, Phys. Rev. B 72 (2005)
	140408.
	\newblock \href {https://doi.org/10.1103/PhysRevB.72.140408}
	{\path{doi:10.1103/PhysRevB.72.140408}}.
	\newline\urlprefix\url{https://link.aps.org/doi/10.1103/PhysRevB.72.140408}

	\bibitem{0305-4470-36-14-101}
	I.~Peschel, \href{http://stacks.iop.org/0305-4470/36/i=14/a=101}{Calculation of
		reduced density matrices from correlation functions}, Journal of Physics A:
	Mathematical and General 36~(14) (2003) L205.
	\newline\urlprefix\url{http://stacks.iop.org/0305-4470/36/i=14/a=101}

	\bibitem{PhysRevB.69.075111}
	S.-A. Cheong, C.~L. Henley,
	\href{https://link.aps.org/doi/10.1103/PhysRevB.69.075111}{Many-body density
		matrices for free fermions}, Phys. Rev. B 69 (2004) 075111.
	\newblock \href {https://doi.org/10.1103/PhysRevB.69.075111}
	{\path{doi:10.1103/PhysRevB.69.075111}}.
	\newline\urlprefix\url{https://link.aps.org/doi/10.1103/PhysRevB.69.075111}

	\bibitem{PhysRevB.22.1305}
	C.~Dasgupta, S.-k. Ma,
	\href{https://link.aps.org/doi/10.1103/PhysRevB.22.1305}{Low-temperature
		properties of the random heisenberg antiferromagnetic chain}, Phys. Rev. B 22
	(1980) 1305--1319.
	\newblock \href {https://doi.org/10.1103/PhysRevB.22.1305}
	{\path{doi:10.1103/PhysRevB.22.1305}}.
	\newline\urlprefix\url{https://link.aps.org/doi/10.1103/PhysRevB.22.1305}

	\bibitem{PhysRevB.102.014455}
	X.~Turkeshi, P.~Ruggiero, V.~Alba, P.~Calabrese,
	\href{https://link.aps.org/doi/10.1103/PhysRevB.102.014455}{Entanglement
		equipartition in critical random spin chains}, Phys. Rev. B 102 (2020)
	014455.
	\newblock \href {https://doi.org/10.1103/PhysRevB.102.014455}
	{\path{doi:10.1103/PhysRevB.102.014455}}.
	\newline\urlprefix\url{https://link.aps.org/doi/10.1103/PhysRevB.102.014455}

	\bibitem{Igloi2018}
	F.~Igl{\'o}i, C.~Monthus,
	\href{https://doi.org/10.1140/epjb/e2018-90434-8}{Strong disorder rg approach
		-- a short review of recent developments}, The European Physical Journal B
	91~(11) (2018) 290.
	\newblock \href {https://doi.org/10.1140/epjb/e2018-90434-8}
	{\path{doi:10.1140/epjb/e2018-90434-8}}.
	\newline\urlprefix\url{https://doi.org/10.1140/epjb/e2018-90434-8}

	\bibitem{PhysRevB.50.3799}
	D.~S. Fisher, \href{https://link.aps.org/doi/10.1103/PhysRevB.50.3799}{Random
		antiferromagnetic quantum spin chains}, Phys. Rev. B 50 (1994) 3799--3821.
	\newblock \href {https://doi.org/10.1103/PhysRevB.50.3799}
	{\path{doi:10.1103/PhysRevB.50.3799}}.
	\newline\urlprefix\url{https://link.aps.org/doi/10.1103/PhysRevB.50.3799}

	\bibitem{PhysRevX.4.011052}
	D.~Pekker, G.~Refael, E.~Altman, E.~Demler, V.~Oganesyan,
	\href{https://link.aps.org/doi/10.1103/PhysRevX.4.011052}{Hilbert-glass
		transition: New universality of temperature-tuned many-body dynamical quantum
		criticality}, Phys. Rev. X 4 (2014) 011052.
	\newblock \href {https://doi.org/10.1103/PhysRevX.4.011052}
	{\path{doi:10.1103/PhysRevX.4.011052}}.
	\newline\urlprefix\url{https://link.aps.org/doi/10.1103/PhysRevX.4.011052}

	\bibitem{PhysRevLett.93.260602}
	G.~Refael, J.~E. Moore,
	\href{https://link.aps.org/doi/10.1103/PhysRevLett.93.260602}{Entanglement
		entropy of random quantum critical points in one dimension}, Phys. Rev. Lett.
	93 (2004) 260602.
	\newblock \href {https://doi.org/10.1103/PhysRevLett.93.260602}
	{\path{doi:10.1103/PhysRevLett.93.260602}}.
	\newline\urlprefix\url{https://link.aps.org/doi/10.1103/PhysRevLett.93.260602}

	\bibitem{PhysRevB.88.075123}
	M.~Pouranvari, K.~Yang,
	\href{https://link.aps.org/doi/10.1103/PhysRevB.88.075123}{Entanglement
		spectrum and entangled modes of random $xx$ spin chains}, Phys. Rev. B 88
	(2013) 075123.
	\newblock \href {https://doi.org/10.1103/PhysRevB.88.075123}
	{\path{doi:10.1103/PhysRevB.88.075123}}.
	\newline\urlprefix\url{https://link.aps.org/doi/10.1103/PhysRevB.88.075123}

	\bibitem{PhysRevB.92.245134}
	M.~Pouranvari, K.~Yang,
	\href{https://link.aps.org/doi/10.1103/PhysRevB.92.245134}{Entanglement
		spectrum and entangled modes of highly excited states in random $xx$ spin
		chains}, Phys. Rev. B 92 (2015) 245134.
	\newblock \href {https://doi.org/10.1103/PhysRevB.92.245134}
	{\path{doi:10.1103/PhysRevB.92.245134}}.
	\newline\urlprefix\url{https://link.aps.org/doi/10.1103/PhysRevB.92.245134}

	\bibitem{https://doi.org/10.48550/arxiv.2208.11409}
	M.~Pouranvari, \href{https://arxiv.org/abs/2208.11409}{Entanglement properties
		of disordered free fermion systems with random bi-partitioning} (2022).
	\newblock \href {https://doi.org/10.48550/ARXIV.2208.11409}
	{\path{doi:10.48550/ARXIV.2208.11409}}.
	\newline\urlprefix\url{https://arxiv.org/abs/2208.11409}

\end{thebibliography}
\end{document}